\documentstyle[preprint,aps,psfig,amssymb]{revtex} 
\tightenlines
\begin{document}
\title{Thermodynamics of hot dense H-plasmas:
Path integral Monte Carlo simulations and analytical approximations}
\author{V.S.~Filinov$^{1}$\footnote{Mercator guest professor at Rostock University},
M.~Bonitz$^2$, W.~Ebeling$^3$, and V.E.~Fortov$^1$}
\address{$^1$Russian Academy of Sciences,
High Energy Density Research Center,
Izhorskaya street 13-19, Moscow 127412, Russia}
\address{$^2$Fachbereich Physik, Universit{\"a}t Rostock\\
Universit{\"a}tsplatz 3, D-18051 Rostock, Germany}
\address{$^3$Institut f\"ur Physik, Humboldt-Universit{\"a}t Berlin\\
Invalidenstrasse 110 D-10115 Berlin}
\date{\today}
\maketitle
\begin{abstract}
This work is devoted to the thermodynamics of high-temperature dense
hydrogen plasmas in the pressure region between $10^{-1}$ and $10^2$ Mbar.
In particular we present for this region results of extensive
calculations based on a
recently developed path integral Monte Carlo scheme (direct PIMC).
This method allows for a correct treatment of the
thermodynamic properties of hot dense Coulomb systems.
Calculations were performed in a broad region of the nonideality parameter
$\Gamma \lesssim 3$ and degeneracy parameter $n_e \Lambda^3 \lesssim 10$. We give a
comparison with a few available results from other
path integral calculations (restricted PIMC) and with analytical
calculations based on Pad\'e approximations for strongly ionized plasmas.
Good agreement between the results obtained from the three independent
methods is found.
\end{abstract}
\pacs{  }
\section{Introduction}

The thermodynamics of strongly correlated Fermi systems at high pressures
are of growing importance in many fields, including shock and laser plasmas,
astrophysics, solids and nuclear matter, see
Refs.~\cite{boston97,binz96,green-book,kbt99} for an overview.
In particular the thermodynamical properties of hot dense plasmas
under high pressure are of importace for the description
of plasmas relevant for laser fusion \cite{ebfhr96}.
Further among the phenomena of current interest are Fermi liquids,
metallic hydrogen \cite{dasilva-etal.97}, plasma phase transition, e.g. 
\cite{sbt95} and references therein,
bound states etc., which occur in situations where both
Coulomb {\em and} quantum effects are relevant. There has been significant
progress in recent years to study these systems analytically and numerically, see e.g.
\cite{sbt95,egger99,EbSt99,KTR94,EbSc97,bonitz-etal.96jpcm,bonitz-book}. 
Due to the enormeous difficulties to
develop analytical descriptions for hydrogen plasmas with strong
coupling, e.g. \cite{boston97,binz96,green-book}, there is still an urgent need
to test the analytical theory by an independent numerical approach.

An approach which is particularly well suited to describe thermodynamic
properties in the region of high pressure, characterized by
strong coupling and strong degeneracy,
is the path integral quantum Monte Carlo (PIMC) method. There has been
remarkable recent progress in applying these techniques to  Fermi systems,
for an overview see e.g. Refs.~\cite{boston97,binz96,zamalin,binder96,berne98}.
However, these simulations are essentially hampered by the fermion sign
problem.
To overcome this difficulty, several strategies have been developed to
simulate macroscopic Coulomb systems \cite{egger99,ceperley95,ceperley95rmp}: the
first is the restricted PIMC concept where additional assumptions on the
density operator ${\hat \rho}$ are introduced which reduce the sum over
permutations to even (positive) contributions only. This requires knowledge
of the nodes of the density matrix which is available only in a few special
cases, e.g. \cite{ceperley95,ceperley95rmp}. However, for interacting
macroscopic systems, these nodes are known only approximately, e.g.
\cite{mil-pol}, and the accuracy of the results is difficult to assess from
within this scheme.
An alternative are direct fermionic PIMC simulations which have occasionally
been attempted by various groups \cite{imada84} but which were not
sufficiently precise and efficient for practical purposes.
Recently, three of us have proposed a new path integral representation
for the N-particle density operator
\cite{filinov-etal.KBT,filinov-etal.99xxx1}
which  allows for {\em direct fermionic path integral Monte Carlo} simulations
of dense plasmas in a broad range of densities and temperatures. Using this
concept we computed the pressure and energy of a degenerate strongly coupled
hydrogen plasma
\cite{filinov-etal.99xxx1,filinov-etal.00jetpl} and the
pair distribution functions in the region of partial ionization and
dissociation \cite{filinov-etal.00jetpl,filinov-etal.00jp}. This scheme is
rather efficient when the number of time slices (beads) in the path integral
is less or equal 50 and was found to
work well for temperatures $k_BT \gtrsim 0.1Ry$.
In this paper we derive further improved 
formulas for the pressure and energy and give, for the first time,
a detailed derivation of all main results and rigorously justify the use of the 
effective quantum pair potential (Kelbg potential) in direct PIMC simulations.
Further, in the present work this method will be applied
to high-pressure plasmas ($p \simeq 10^{-1} - 10^2 $Mbar) in such temperature
regions were considerable deviations from the ideal
behavior are observed.

One difficulty of PIMC simulations is that reliable error estimates are
often not available, in particular for strongly coupled degenerate systems.
Moreover, in this region no reliable data from other theories such as density
functional theory or quantum statistics, e.g. \cite{green-book,bonitz-book}, 
are available
which would allow for an unambiguous test. Furthermore, results from classical 
molecular dynamics simulations exist, but they apply only to fully ionized 
and weakly degenerate plasmas, e.g. \cite{zwick_pr,vova1,vova2}, which is outside the 
range of interest for this work. Also, new quantum molecular dynamics approaches are 
being developed, e.g. \cite{KTR94,zwick_sm,filinov_qmd}, but they are only beginning to 
produce accurate results. 

Therefore, it is of high interest
to perform quantitative comparisons of independent simulations, such as
restricted and direct fermionic PIMC, and to develop improved analytical
approximations, which is the aim of this paper. We
compare recent results of Militzer et al.
\cite{militzer-etal.00} for pressure and energy isochors
($n\sim 2.5 \cdot  10^{23} {\rm cm}^{-3}$) of dense
hydrogen to our own direct PIMC results. This is a non-trivial comparison
since the two approaches employ independent
sets of approximations. Nevertheless, we find very good agreement
for temperatures ranging from $10^6 K$ to as low as $50,000K$.
This is remarkable since there the coupling and degeneracy parameters reach
rather large values, $\Gamma \approx 3$ and $n_e \Lambda^3 \approx 10$,
and the plasma contains a substantial fraction of bound states.

Further, we use the new data to make a comparison with analytical estimates which are 
based on
 Pad\'e approximations for strongly ionized plasmas. These formulae were
constructed on the basis of the known analytical results
for the limiting cases of low density \cite{green-book,ebeling} and
high density \cite{green-book}. These Pad\'e approximations are exact up to
quadratic terms in the density and interpolate between
the virial expansions and the high-density asymptotics
\cite{Pade85,Pade90,EbFo91}. We find that the results for the internal energy
and for the pressure agree well with the PIMC results
in the region of the density temperature plane, where 
$\Gamma \lesssim 1.6 $ and $n\Lambda^3 \lesssim 5$.

\section{Path integral representation of thermodynamic quantities}
We now our direct PIMC scheme.
All thermodynamic properties of a two-component plasma are defined
by the partition function $Z$ which, for the case of $N_e$
electrons and $N_p$ protons, is given by
\begin{eqnarray}
Z(N_e,N_p,V,\beta) &=&
\frac{Q(N_e,N_p,\beta)}{N_e!N_p!},
\nonumber\\
\mbox{with} \qquad
Q(N_e,N_p,\beta) &=& \sum_{\sigma}\int\limits_V dq \,dr
\,\rho(q, r,\sigma;\beta),
\label{q-def}
\end{eqnarray}
where $\beta=1/k_B T$.
The exact density matrix is, for a quantum system, in
general, not known but can be constructed using a path integral representation
\cite{feynman-hibbs},
\begin{eqnarray}
\int\limits_{V} dR^{(0)}\sum_{\sigma}\rho(R^{(0)},\sigma;\beta) &=&
 \int\limits_{V} dR^{(0)} \dots dR^{(n)} \,
\rho^{(1)}\cdot\rho^{(2)} \, \dots \rho^{(n)}
\nonumber\\
&\times&\sum_{\sigma}\sum_{P} (\pm 1)^{\kappa_P}
\,{\cal S}(\sigma, {\hat P} \sigma')\,
{\hat P} \rho^{(n+1)},
\label{rho-pimc}
\end{eqnarray}
where $\rho^{(i)}\equiv \rho\left(R^{(i-1)},R^{(i)};\Delta\beta\right) \equiv
\langle R^{(i-1)}|e^{-\Delta \beta {\hat H}}|R^{(i)}\rangle$,
whereas $\Delta \beta \equiv \beta/(n+1)$ and
$\Delta\lambda_a^2=2\pi\hbar^2 \Delta\beta/m_a$, $a=p,e$.
${\hat H}$ is the Hamilton operator, ${\hat H}={\hat K}+{\hat U}_c$, containing
kinetic and potential energy contributions, ${\hat K}$ and ${\hat U}_c$,
respectively, with
${\hat  U}_c = {\hat  U}_c^p + {\hat  U}_c^e + {\hat  U}_c^{ep}$ being the sum
of the Coulomb potentials between protons (p), electrons (e) and electrons and
protons (ep)].
Further,
$R^{(i)}=(q^{(i)},r^{(i)}) \equiv (R_p^{(i)},R_e^{(i)})$, for $i=1,\dots n+1$,
$R^{(0)}\equiv (q,r)\equiv (R_p^{(0)},R_e^{(0)})$, and
$R^{(n+1)} \equiv R^{(0)}$ and $\sigma'=\sigma$. This means, the particles are
represented by fermionic loops with the coordinates (beads)
$[R]\equiv [R^{(0)}; R^{(1)};\dots; R^{(n)}; R^{(n+1)}]$, where $q$ and
$r$ denote the electron and proton coordinates, respectively.
The spin gives rise to the spin part of the density matrix ${\cal S}$, whereas
exchange effects are accounted for by the permutation operator ${\hat P}$, which
acts on the electron coordinates and spin projections, and
the sum over the permutations with parity $\kappa_P$. In the fermionic case
(minus sign), the sum contains $N_e!/2$ positive and negative terms leading
to the notorious sign problem. Due to the large mass difference of electrons
and ions, the exchange of the latter is not included.
The matrix elements $\rho^{(i)}$ can be rewritten identically as
\begin{eqnarray}
&&\langle R^{(i-1)}|e^{-\Delta \beta {\hat H}}|R^{(i)}\rangle =
\nonumber\\
&&\quad\int d{\tilde p}^{(i)}d{\bar p}^{(i)}\,
\langle R^{(i-1)}|e^{-\Delta \beta {\hat U}_c}|{\tilde p}^{(i)}\rangle
\langle {\tilde p}^{(i)}|e^{-\Delta \beta {\hat K}}
|{\bar p}^{(i)}\rangle
\langle {\bar p}^{(i)}|
\,e^{-\frac{\Delta \beta^2}{2}[{\hat K},{\hat U}_c]} \, \dots|R^{(i)}\rangle.
\label{rho_ku}
\end{eqnarray}

To compute thermodynamic functions, the logarithm of the partition function
has to be differentiated with respect to thermodynamic variables. In particular,
for the equation of state $p$ and internal
energy $E$ follows,
\begin{eqnarray}
\beta p &=& \partial {\rm ln} Q / \partial V =
[\alpha/3V \partial {\rm ln} Q
/ \partial \alpha]_{\alpha=1},
\label{p_gen}
\\
\beta E &=& -\beta \partial {\rm ln} Q
/ \partial \beta,
\label{e_gen}
\end{eqnarray}
where $\alpha$ is a length scaling parameter $\alpha = L/L_0$. This means,
in the path integral representation (\ref{rho-pimc}), each high-temperature
density matrix has to be differentiated in turn. For example, the result for
the energy will have the form
\begin{eqnarray}
\beta E &=& - \frac{1}{Q}
\int\limits_{V} dR^{(0)} \dots dR^{(n)} \,
\nonumber\\
&\times & \sum_{k=1}^{n+1}\rho^{(1)}\dots\rho^{(k-1)}\cdot
\left[\beta\frac{\partial \rho^{(k)}}{\partial \beta}\right]
\cdot \rho^{(k+1)} \, \dots \,\rho^{(n)}
\sum_{\sigma}\sum_{P}(\pm 1)^{\kappa_P}
\,{\cal S}(\sigma, {\hat P} \sigma'){\hat P} \rho^{(n+1)},
\label{e-pimc}
\end{eqnarray}
and, analogously for other thermodynamic functions.

There are two different approaches to evaluate this expression. One is to
first choose an approximation for the high-temperature density matrices
$\rho^{(i)}$ and then to perform the differentiation. The other way is to
first differentiate the operator expression for $\rho^{(k)}$ and
use an approximation for the matrix elements only in the final result.
As we checked, the second method is
more accurate and will be used in the following.

To evaluate the derivatives
in Eq. (\ref{e-pimc}), it is convenient to indroduce dimensionless
integration variables $\eta^{(k)}=(\eta_p^{(k)},\eta_e^{(k)})$, where
$\eta_a^{(k)}=\kappa_a(R^{(k)}_a-R^{(k-1)}_a)$ for $k=1,\dots, n$ and
$a=p,e$, and $\kappa_a^2\equiv m_a k_BT/(2\pi \hbar^2)=1/\lambda_a^2$,
\cite{filinov-etal.99xxx1}. This has the advantage that now the differentiation
of the density matrix affects only the interaction terms. Indeed, one can show that
\begin{eqnarray}
\beta \frac{\partial \rho^{(k)}}{\partial \beta} &=&
-\beta \frac{\partial \Delta\beta \cdot U_c(X^{(k-1)})}{\partial \beta}
\rho^{(k)} +\beta {\tilde \rho}_{\beta}^{(k)},
\label{rho-prime}
\end{eqnarray}
where $X^{(0)}\equiv (\kappa_p R_p^{(0)},\kappa_e R_e^{(0)})$,
$X^{(k)}\equiv (X_p^{(k)},X_e^{(k)})$, with
$X_a^{(k)}=\kappa_a R_a^{(0)}+\sum_{l=1}^k \eta^{(l)}_a$, and $k$ runs from $1$ to $n$.
Further, $X^{(n+1)}\equiv (\kappa_p R_p^{(n+1)},\kappa_e R_e^{(n+1)})=X^{(0)}$,
and we denoted
\begin{eqnarray}
{\tilde \rho}_{\beta}^{(k)} =
\int d{p}^{(k)}\,
\langle X^{(k-1)}|e^{-\Delta \beta {\hat U}_c}|{p}^{(k)}\rangle
e^{-\frac{\langle {p}^{(k)}|{p}^{(k)}\rangle}{4\pi(n+1)}}
\langle {p}^{(k)}|\frac{\partial}{\partial \beta}
\,e^{-\frac{(\Delta\beta)^2}{2}[{\hat K},{\hat U}_c]} \, \dots|X^{(k)}\rangle,
\label{rho-tilde}
\end{eqnarray}
where $p_a^{(k)}={\tilde p}_a^{(k)}/(\kappa_a\hbar)$, $p{(k)}\equiv (p_p^{(k)},p_e^{(k)})$
and use has been made of Eq. (\ref{rho_ku}).
For $k=n+1$, we have
\begin{eqnarray}
&&\beta \frac{\partial}{\partial \beta}
\sum_{\sigma}\sum_{P}(\pm 1)^{\kappa_P}
\,{\cal S}(\sigma, {\hat P} \sigma'){\hat P} \rho^{(n+1)}=
\sum_{\sigma}\sum_{P}(\pm 1)^{\kappa_P}
\,{\cal S}(\sigma, {\hat P} \sigma')\times
\nonumber\\
&\times&
\left\{ -\beta \frac{\partial \Delta\beta \cdot U_c(X^{(n)})}{\partial \beta}
{\hat P}\rho^{(n+1)}
+ {\hat P}\left[ \beta {\tilde \rho}_{\beta}^{(n+1)}
 \right]
 \right\}.
\label{rho-prime1}
\end{eqnarray}
Further, $U_c(X^{(k-1)})\equiv U^{(1)}_c(X^{(k-1)})+U^{(2)}_c(X^{(k-1)})$, with
$U_c^{(1)}$ and $U_c^{(2)}$ denoting the interaction between identical and different particle
species, respectively,
$U^{(1)}_c(X)=U^e_c(X)+U^p_c(X)$ and $U^{(2)}_c(X)=U^{ep}_c(X)$.

Using these results and Eq. (\ref{e-pimc}), we obtain for the energy
\begin{eqnarray}
\beta E &=& \frac{3}{2}(N_e+N_p) -
 \frac{1}{Q}
\frac{1}{\,\lambda_p^{3N_p}\lambda_e^{3N_e}}
\int\limits_{V} dR^{(0)} d\eta^{(1)} \dots d\eta^{(n)} \,
\sum_{\sigma}\sum_{P}(\pm 1)^{\kappa_P}
\,{\cal S}(\sigma, {\hat P} \sigma')
\nonumber\\
&\times & \Bigg\{\sum_{k=1}^{n+1}\rho^{(1)}\dots\rho^{(k-1)}
\left[
-\beta \frac{\partial \Delta\beta \cdot U^{(1)}_c(X^{(k-1)})}{\partial \beta}
-\beta \frac{\partial \Delta\beta \cdot U^{(2)}_c(X^{(k-1)})}{\partial \beta}
+\beta {\tilde \rho}_{\beta}^{(n+1)} \right]
\nonumber\\
&\times & \rho^{(k)} \, \dots \,\rho^{(n)}{\hat P} \rho^{(n+1)}
\Bigg\}\bigg|_{X^{(n+1)}=X^{(0)},\, \sigma'=\sigma}.
\label{pimc1}
\end{eqnarray}
This way, the derivative of the density matrix has been calculated,
and we turn to the next point - to find approximations for the high-temperature
density matrix.

\section{High-temperature asymptotics of the
density matrix in the Path integral approach. Kelbg potential}
In this section we derive an approximation for the high-tempature 
density matrix which is suitable for direct PIMC simulations. Further, we 
demonstrate that the proper choice of the effective quantum pair potential is 
given by the Kelbg potential.
Following Refs. \cite{zamalin,filinov76,znf76},
we derive a modified representation for the density matrix. The mains steps 
are:
\begin{enumerate}

\item{The N-particle density matrix is expanded in terms of
2-particle, 3-particle etc. contributions from which only the first,
$\rho_{ab}$, is retained \cite{zamalin,filinov76,znf76};}

\item{In the high-temperature limit, $\rho_{ab}$ factorizes into a kinetic
($\rho_0$) and an interaction term ($\rho_U^{ab}$),
$\rho_{ab}\approx \rho_0\rho_U^{ab}$,
 because it can be shown that  \cite{Ke63,kelbg}
\begin{eqnarray}
e^{-\frac{(\Delta\beta)^2}{2}[{\hat K},{\hat U}_c]} =
{\hat I} + O\left(\frac{1}{(n+1)^2}\right),
\end{eqnarray}
where ${\hat I}$ is the unity operator.
In this way we get the following representation for the two-particle
density matrix
\begin{equation}
\rho_{ab} = \left(\frac{(m_a m_b)^{3/2}}{(2 \pi \hbar \beta)^3}\right)
\exp[-\frac{m_a}{2 \hbar^2 \beta} ({\bf r}_a - {\bf r}_a')^2]
\exp[-\frac{m_b}{2 \hbar^2 \beta} ({\bf r}_b - {\bf r}_b')^2]
\exp[-\beta \Phi_{ab}]
\end{equation}
where $\Phi_{ab}({\bf r}_a,{\bf r}_a', {\bf r}_b, {\bf r}_b')$ is the
off-diagonal two-particle effective potential.}

\item{In the following, the off-diagonal matrix elements of the effective
binary potentials will be approximated
by the diagonal ones by taking the Kelbg potential at the center coordinate,
$\Phi^{ab}(r,r';\Delta\beta)\approx \Phi^{ab}(\frac{r+r'}{2};\Delta\beta)$.;}
\item{For the plasma parameter region of interest, the protons can be treated
classically, and $\Phi^{ii}$ may be approximated by the Coulomb potential.}
\end{enumerate}

We will now comment on these steps in some more detail.
We calculated the effective potential by
solving a Bloch equation by first order perturbation theory. The method has been described in
detail in \cite{kelbg}.
This procedure defines an effective off-diagonal quantum pair
potential for Coulomb systems,
which depends on the inter-particle distances
${\bf r}_{ab},{\bf r}'_{ab}$. As a result of first-order perturbation theory
we get explicitely

\begin{eqnarray}
\Phi^{ab}({\bf r}_{ab},{\bf r}'_{ab},\Delta\beta) \equiv
e_a e_b \,\int_{0}^{1}\frac{d \alpha}{d_{ab}(\alpha)}
{\rm erf}\left(\frac{d_{ab}(\alpha)}{2\lambda_{ab}\sqrt{\alpha(1-\alpha)}}\right),
\label{kelbg-off}
\end{eqnarray}
where $d_{ab}(\alpha)=|\alpha {\bf r}_{ab}+(1-\alpha){\bf r}'_{ab}|$,
${\rm erf}(x)$ is the error function
${\rm erf}(x)=\frac{2}{\sqrt{\pi}}\int_0^x dt e^{-t^2}$, and
$\lambda^2_{ab}=\frac{\hbar^2\Delta\beta}{2\mu_{ab}}$ with
$\mu_{ab}^{-1}=m_a^{-1}+m_b^{-1}$. It is interesting to note, that a
simple approximation of the complicated integral over $\alpha$ by
the length of the interval multiplied with the integrand in the center
(Mittelwertsatz) leads us to the so-called KTR-potential due to Klakow, Toepffer
and Reinhard which (in the diagonal
approximation) is often used
in quasi-classical MD simulations \cite{KTR94,EbSc97}

\begin{eqnarray}
\Phi^{ab}({\bf r}_{ab},{\bf r}'_{ab},\Delta\beta) \equiv
\frac{e_a e_b}{d_{ab}(1/2)}
{\rm erf}\left(\frac{d_{ab}(1/2)}{\lambda_{ab}}\right),
\label{KTR-pot}
\end{eqnarray}

In our direct PIMC calculations we used the full expression for
the interaction potential, keeping the
$\alpha$-integration but, in order to save computer time, we
approximated the two-particle interaction potential by its diagonal elements.
The diagonal element
(${\bf r}'_{ab}={\bf r}_{ab}$) of $\Phi^{ab}$ is just the familiar Kelbg
potential, given by (we will use the same notation for the potential)

\begin{eqnarray}
\Phi^{ab}(|{\bf r}_{ab}|,\Delta\beta)\equiv
\Phi^{ab}({\bf r}_{ab},{\bf r}_{ab},\Delta\beta) =
\frac{e_a e_b}{\lambda_{ab} x_{ab}} \,\left[1-e^{-x_{ab}^2}+\sqrt\{\pi\} x_{ab}
\left(1-{\rm erf}(x_{ab})\right)
\right],
\label{kelbg-d}
\end{eqnarray}
where $x_{ab}=|{\bf r}_{ab}|/\lambda_{ab}$, and we underline that the Kelbg
potential is finite at zero distance.
The error of the above approximations, for each of the high-temperature
factors on the r.h.s. of Eq. (\ref{rho-pimc}), is of the order
$1/(n+1)^2$.

With these approximations, we obtain the result
$\rho^{(k)}=\rho_0^{(k)}\rho_U^{(k)}+O[(1/n+1)^2]$,
where $\rho_0^{(k)}$ is the kinetic density matrix, while
$\rho_U^{(k)}=e^{-\Delta \beta U(X^{(k-1)})}\delta(X^{(k-1)}-X^{(k)})$,
where $U$ denotes the following sum of Coulomb and Kelbg potentials,
$U(X^{(k)})=U_c^p(X_p^{(k)})+U^e(X_e^{(k)})+U^{ep}(X_p^{(k)},X_e^{(k)})$.
Notice that special care has to be taken in performing the derivatives with
respect to $\beta$ of the Coulomb potentials which appear in Eq. (\ref{pimc1}).
Indeed, products $\rho^{(1)} \, \dots \,\rho^{(n)}{\hat P} \rho^{(n+1)}
\beta \frac{\partial \Delta\beta \cdot U_c(X^{(k-1)})}{\partial \beta}$ have
a singularity at zero interparticle distance which is integrable but
leads to difficulties in the simulations. Due to the Kelbg potential, for the
e-e and p-p interaction, this singularity is weakenend, but it is enhanced for
the e-p interaction. In order to assure efficient simulations we, therefore,
further transform the e-p contribution in the following way:
\begin{eqnarray}
&&\int_0^1  d\alpha \int dR^{(k-1)}
\langle R^{(k-2)}|e^{-\Delta \beta \alpha{\hat K}} |R^{(k-1)}\rangle
\left[-\beta\frac{\partial}{\partial\beta}
\left(\Delta \beta U_c^{(2)}(R^{(k-1)})\right)
\right]
\nonumber\\
&\qquad& \times
\langle R^{(k-1)}|e^{-\Delta \beta (1-\alpha){\hat K}} |R^{(k)}\rangle
\nonumber\\
&\approx &
\langle R^{(k-1)}|e^{-\Delta \beta {\hat K}} |R^{(k)}\rangle
\left[-\beta\frac{\partial}{\partial\beta}
\left(\Delta \beta U^{(2)}(R^{(k-1)})\right)
\right] + O\left[(1/n+1)^2\right].
\label{uep}
\end{eqnarray}
This means, within the standard error of our approximation
$O\left[(\Delta \beta)^2\right]$, we have replaced the e-p Coulomb potential
$U_c^{(2)}$ by the corresponding Kelbg potential $U^{(2)}$, which is
much better suited for MC simulations.

Thus, using $\lambda_p \ll \lambda_e$, we finally obtain for the energy:

\begin{eqnarray}
\beta E = \frac{3}{2}(N_e+N_p) + \frac{1}{Q}
\frac{1}{\,\lambda_p^{3N_p}\Delta \lambda_e^{3N_e}}\sum_{s=0}^{N_e}
\int dq \, dr \, d\xi \,\rho_s(q,[r],\beta) \,\times
\nonumber\\
\Bigg\{\sum_{p<t}^{N_p} \frac{\beta e^2}{|q_{pt}|} +
\sum_{l=0}^{n}\Bigg[\sum_{p<t}^{N_e} \frac{\Delta\beta e^2}{|r^l_{pt}|}
+  \sum_{p=1}^{N_p}\sum_{t=1}^{N_e} \Psi_l^{ep}\Bigg]
\nonumber\\
+ \sum_{l=1}^{n}\Bigg[
- \sum_{p<t}^{N_e}C^l_{pt}
\frac{\Delta\beta e^2}{|r^l_{pt}|^2} +
 \sum_{p=1}^{N_p}\sum_{t=1}^{N_e}
D^l_{pt}
\frac{\partial \Delta\beta\Phi^{ep}}{\partial |x^l_{pt}|}
 \Bigg]
\nonumber\\
\,-\,
\frac{1}{{\rm det} |\psi^{n,1}_{ab}|_s}
\frac{\partial{\rm \,det} | \psi^{n,1}_{ab} |_s}{\partial \beta}
\Bigg\},
\nonumber \\
{\rm with} \quad C^l_{pt} = \frac{\langle r^l_{pt}|y^l_{pt}\rangle}{2|r^l_{pt}|},
\qquad D^l_{pt} = \frac{\langle x^l_{pt}|y^l_{p}\rangle}{2|x^l_{pt}|},
\label{energy}
\quad
\end{eqnarray}
and $\Psi_l^{ep}\equiv \Delta\beta\partial
[\beta'\Phi^{ep}(|x^l_{pt}|,\beta')]/\partial\beta'|_{\beta'=\Delta\beta}$
contains the electron-proton Kelbg potential $\Phi^{ep}$.
Here,
$\langle \dots | \dots \rangle$ denotes the scalar product, and
$q_{pt}$, $r_{pt}$ and $x_{pt}$ are differences of two
coordinate vectors:
$q_{pt}\equiv q_p-q_t$,
$r_{pt}\equiv r_{p}-r_{t}$, $x_{pt}\equiv r_p-q_t$, $r^l_{pt}=r_{pt}+y_{pt}^l$,
 $x^l_{pt}\equiv x_{pt}+y^l_p$ and
 $y^l_{pt}\equiv y^l_{p}-y^l_{t}$, with $y_a^n=\Delta\lambda_e\sum_{k=1}^{n}\xi^{(k)}_a$.
Here we introduced dimensionless distances between neighboring
vertices on the loop, $\xi^{(1)}, \dots \xi^{(n)}$,
thus, explicitly,
$[r]\equiv [r; y_e^{(1)}; y_e^{(2)}; \dots].$ Further,
the density matrix $\rho_s$ in Eq. (\ref{energy}) is given by
\begin{eqnarray}
\rho_s(q,[r],\beta) = C^s_{N_e}
\, e^{-\beta U(q,[r],\beta)} \prod\limits_{l=1}^n
\prod\limits_{p=1}^{N_e} \phi^l_{pp}
{\rm det} \,|\psi^{n,1}_{ab}|_s,
\label{rho_s}
\end{eqnarray}
where
$U(q,[r],\beta)=
U_c^p(q)+\{U^e([r],\Delta\beta)+U^{ep}(q,[r],\Delta\beta)\}/(n+1)$
and  $\phi^l_{pp}\equiv \exp[-\pi |\xi^{(l)}_p|^2]$.
We underline that the density matrix (\ref{rho_s})
does not contain an explicit
sum over the permutations and thus no sum of terms with alternating
sign. Instead, the whole exchange problem is
contained in a single exchange matrix given by
\begin{eqnarray}
||\psi^{n,1}_{ab}||_s\equiv ||e^{-\frac{\pi}{\Delta\lambda_e^2}
\left|(r_a-r_b)+ y_a^n\right|^2}||_s.
\label{psi}
\end{eqnarray}
As a result of the spin summation,
the matrix carries a subscript $s$ denoting the number of electrons having
the same spin projection. For more details, we refer to
Refs. \cite{filinov-etal.KBT,filinov-etal.99xxx1}.

In similar way, we obtain the result for the equation of state,
\begin{eqnarray}
\frac{\beta p V}{N_e+N_p} = 1 + \frac{1}{N_e+N_p}
\frac{(3Q)^{-1}}{\,\lambda_p^{3N_p}\Delta\lambda_e^{3N_e}}\sum_{s=0}^{N_e}
\int dq \, dr \, d\xi \,\rho_s(q,[r],\beta) \times
\nonumber\\
\Bigg\{\sum_{p<t}^{N_p} \frac{\beta e^2}{|q_{pt}|} +
\sum_{p<t}^{N_e}\frac{\Delta\beta e^2}{|r_{pt}|}
-  \sum_{p=1}^{N_p}\sum_{t=1}^{N_e} |x_{pt}|
\frac{\partial \Delta\beta\Phi^{ep}}{\partial |x_{pt}|}
\nonumber\\
+\sum_{l=1}^{n}\left[\sum_{p<t}^{N_e}
A^l_{pt}
\frac{\Delta\beta e^2}{|r^l_{pt}|^2}
- \sum_{p=1}^{N_p}\sum_{t=1}^{N_e}B^l_{pt}
\frac{\partial \Delta\beta\Phi^{ep}}{\partial |x^l_{pt}|}
\right]
\nonumber\\
\,+\,\frac{\alpha}{{\rm det} |\psi^{n,1}_{ab}|_s}
\frac{\partial{\rm \,det} | \psi^{n,1}_{ab} |_s}{\partial \alpha}
\Bigg\},
\nonumber \\
{\rm with} \quad A^l_{pt} = \frac{\langle r^l_{pt}|r_{pt}\rangle}{|r^l_{pt}|},
\qquad B^l_{pt} = \frac{\langle x^l_{pt}|x_{pt}\rangle}{|x^l_{pt}|}.
\label{eos}
\quad
\end{eqnarray}

The structure of Eqs.~(\ref{energy}, \ref{eos}) is obvious: we have separated the classical
ideal gas part (first term). The ideal quantum part in excess of the
classical one and the correlation
contributions are contained in the
integral term, where the second line results from the ionic correlations
(first term) and the e-e and e-i interaction at the first vertex (second
and third terms respectively). The third and fourth lines are due to the
further electronic vertices and the explicit
temperature dependence [in Eq.~(\ref{energy}) and volume dependence in
Eq.~(\ref{eos})] of the exchange matrix, respectively.
The main advantage of Eqs.~(\ref{energy}, \ref{eos}) is that the
explicit sum over permutations has been converted into the spin determinant
which can be computed very efficiently using standard linear algebra
methods. Furthermore, each of the sums in curly brackets in
Eqs.~(\ref{energy}, \ref{eos}) is bounded as the number of vertices increases,
$n\rightarrow \infty$, and is thus well suited for efficient Monte Carlo
simulations. Notice
also that Eqs.~(\ref{energy}, \ref{eos}) contain the important
limit of an ideal quantum plasma in a natural way \cite{hermann}.

\section{Comparison of direct and restricted PIMC simulations}

Expressions (\ref{energy}, \ref{eos}) are well suited for numerical
evaluation using Monte Carlo techniques, e.g. \cite{zamalin,binder96}.
 In our Monte Carlo scheme
we used three types of steps, where either electron or proton coordinates,
$r_i$ or $q_i$  or
inidividual electronic beads $\xi_i^{(k)}$ were moved until convergence of the
calculated values was reached. Our procedure has been extensively tested.
In particular, we found from comparison with the known analytical expressions
for pressure and energy of an ideal Fermi gas that the Fermi statistics is
very well reproduced \cite{filinov-etal.00jetpl}. Further, we performed
extensive tests for few--electron systems in a harmonic trap where, again,
the analytically known limiting behavior (e.g. energies) is well reproduced
\cite{afilinov-etal.00pss,afilinov-etal.99prl}.
For the present simulations of dense hydrogen, we varied both the particle
number and the number of time slices (beads). As a result of these tests,
we found that to obtain convergent results for the thermodynamic properties
of dense hydrogen, particle numbers $N_e=N_p= 50$ and beads numbers in
the range of $n=6\dots 20$ are adequate
\cite{filinov-etal.99xxx1,filinov-etal.00jetpl}.

\vspace{-2.5cm} 
\begin{figure}[p] 
\centerline{ 
\psfig{file=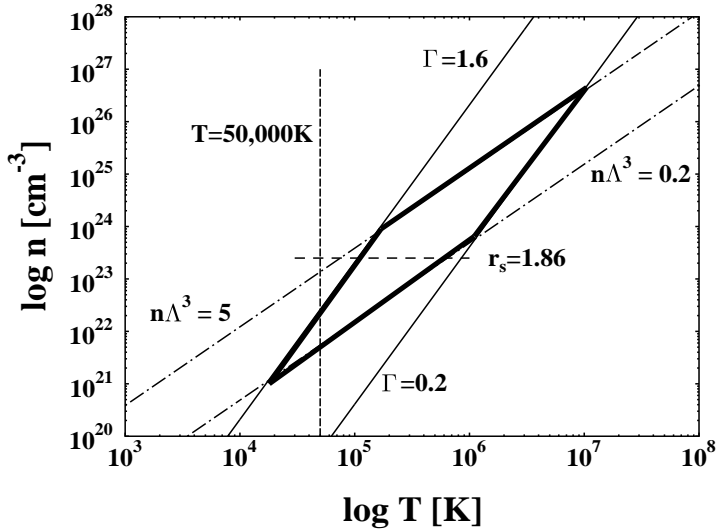,height=18cm,width=13cm}} 
\vspace{-7.5cm} 
\caption[]{\label{n-t} Density-temperature plane showing the parameter  
region for which calculations are performed. The data of Fig.~2 
are along the dashed line (isochor $r_s=1.86$). The data of Figs.~3 and 4 
are inside the bold rhomb, along lines of constant $\Gamma$ between 
the lines $n\Lambda^3=2$ and $n\Lambda^3=5$, respectively. Data for the 
vertical line (isotherm $T=50,000K$) are given in Fig.~5.
} 
\end{figure} 

We will now compare our results with some available results obtained
by the Monte Carlo technique developed by the Urbana group
\cite{ceperley95,militzer-etal.00}.
We may first state that both Monte Carlo techniques differ in several fundamental 
points,
so that they are essentially independent approaches.
Let us briefly outline the main differences between the technique developed
in Urbana, known as the {\em restricted PIMC} scheme
\cite{militzer-etal.00} and references therein, and the approach described here.
These authors performed simulations with
32 electrons and protons; their restricted PIMC scheme required to use a rather small
time step assuring $1 / \Delta \beta \sim 2 * 10^6K$.
Also, the treatment of the interactions differs from our scheme: the authors
of Ref.\cite{militzer-etal.00} perform a numerical solution of the Bloch
equation for the two-particle density matrix whereas we use an analytical
approximation for the effective pair interaction (based on the Kelbg potential,
see above).
Finally, Ref.\cite{militzer-etal.00} approximately computes the nodal surface
of the density matrix using a variational ansatz which is then used to
restrict the integrations to the region of positive density matrix.
For more details regarding the restricted PIMC simulations, see Refs.
\cite{ceperley95,militzer-etal.00}.

\begin{figure}[p] 
\centerline{ 
\psfig{file=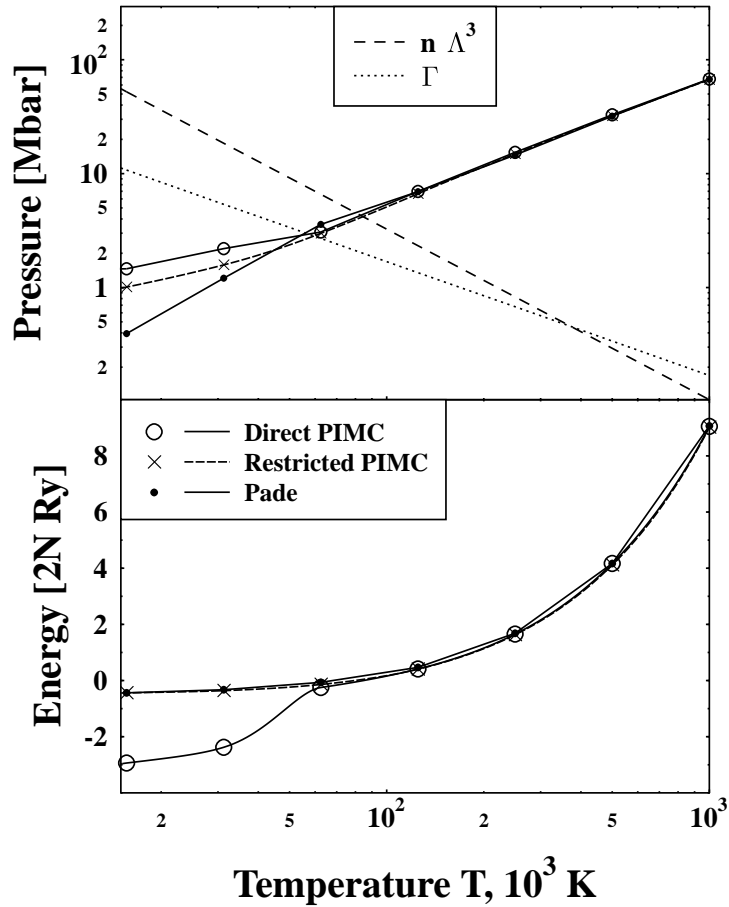,height=15cm}} 
\vspace{-2cm} 
\caption[]{\label{figmil1} Comparison of direct and restricted PIMC results and 
analytical results (PADE)
for the pressure and total energy of dense hydrogen as a function of temperature for 
$r_s=1.86$, corresponding to $n=2.5 \cdot 10^{23}cm^{-3}$. 
For illustration, also the coupling and degeneracy parameters 
$\Gamma$ and $n_e \Lambda^3$ are shown in the upper figure. 
} 
\end{figure} 

Let us now turn to a comparison of the numerical results. The restricted PIMC simulation data
for dense hydrogen are taken from Ref. \cite{militzer-etal.00}. A comparison
of results for the pressure and the internal energy for a fixed value
of the density ($r_s = 1.86$) is shown in Fig. 1 and TABLE I. At high
temperatures, above 50,000 K, where only a small fraction of atoms is
expected, the agreement is rather
good. This is remarkable since the nonideality and the degeneracy
reach values of 3 and 10, respectively. This result demonstrates that, at least 
for $r_s \simeq 1$
and for $T \ge 50,000 K$ both methods yield results which are more or less
equivalent. At $T < 50,000 K$, where partial ionization is expected, we still
observe a reasonable agreement of both approaches, however, we see also
that the differences start to grow. The reasons for that are manyfold. From our 
results we conclude that the main problem is not the bound state formation - 
atoms and molecules are well described by the two PIMC simulations which use 
a physical picture which does not involve any artificial distinction between free 
and bound electrons, e.g. \cite{filinov-etal.00jetpl}. On the other hand,
with growing degeneracy $n\Lambda^3$, both PIMC methods become less reliable, and a 
detailed analysis, although being very desirable, will have to be based on more 
extensive calculations in the future.

Further we present in TAB. I also Pad\'e results
for the weakly nonideal region. We find good agreement with the PIMC results for
$T > 10^5 $ K. Details on the method of these analytical calculations
will be discussed in the next section.

\section{Comparison with analytical approximations for the thermodynamic functions of
strongly ionized dense plasmas}

In this section we  give
a comparison of the available data points from direct PIMC calculations
with analytical estimates based on Pad\'e approximations for strongly ionized
plasmas \cite{green-book,Pade85,Pade90,EbFo91}.
The comparison concentrates on H-plasmas
in a region in the density-temperature plane with the following borders
\begin{eqnarray}
0.2 & \le & \Gamma \le 1.6,
\nonumber\\
0.2 & \le & n_e \Lambda_e^3 \le 5,
\end{eqnarray}
which will be called ``rhomb of moderate nonideality and moderate
degeneracy`` (see bold rhomb in Fig.~1). With respect to analytical treatment,
this rhombic region is
of particular difficulty since none of the known analytic limiting expressions is valid.
Further we calculated several points for $r_s = 1.86$ and $\Gamma \lesssim 2$
which correspond to the PIMC data discussed in the previous section
and also an isotherm at $T = 50,000 K$ including some data at higher density, 
outside the rhomb, cf. Fig.~1 for an overviev.

We demonstrate below that the Pad\'e approximations which interpolate
between the limits where
theoretical results are available are a useful tool for the description of the
available data points, at least for the case of moderate nonideality
$\Gamma \lesssim 1.6$
and moderate degeneracy $n_e \Lambda^3 \lesssim 5$.
The Pad\'e approximations which we use here were constructed in earlier work,
\cite{Pade85,Pade90,EbFo91},
from the known analytical results
for limiting cases of low density \cite{green-book,ebeling} and
high density \cite{green-book}. The structure of the Pad\'e approximations was devised in 
such a way that
they are analytically exact up to
quadratic terms in the density (up to the second virial coefficient) and interpolate 
between
the virial expansions and the high-density asymptotic expressions
\cite{Pade85,Pade90,EbFo91}. The formation of bound states was taken
into account by using a chemical picture.
This means the plasma is considered as a mixture of free electrons, free
ions, atoms and molecules which are in chemical equilibrium, being described
by mass action laws or minimization of the free energy \cite{EbFo91}.

We follow in large here this cited work, only the contribution
of the ion-ion interaction which is, in most cases, the largest one,
was substantially improved following
recent work of Kahlbaum, who succeeded in describing the available
classical Monte Carlo data for the ions by accurate
Pad\'e approximations \cite{Kahlbaum96}.
By using Kahlbaum's formulas we achieve
a rather accurate description of the thermodynamics
in the region where the plasma
behaves like a classical one-component ion plasma imbedded into a sea
of nearly ideal electrons. This is the region where
the electrons are strongly degenerate
\begin{equation}
n_e \Lambda_e^3 \gg 1 \quad\mbox{and}\quad r_s \ll 1,
\end{equation}
and the ions are still classical but nonideal 
\begin{eqnarray}
\Gamma \gg 1 \quad \mbox{and}\quad
n_i \Lambda_i^3 \ll 1.
\end{eqnarray}
This region lies in the upper left corner of Fig.~1.

With respect to the chemical picture we restrict
ourselves to the region of strong ionization where the number of atoms is still 
relatively low
and where the fraction of molecules is small as well, see below.
We will discuss here only the general structure of the Pad\'e formulae. For example, the
internal energy density of the plasma is given by

\begin{equation}
u = u_{id} + u_{int}.
\end{equation}
Here $u_{id}$ is the internal energy of an ideal plasma consisting of Fermi electrons, classical
protons and classical atoms, and $u_{int}$ is the interaction energy
\begin{equation}
u_{int} = u_{ii} + u_{ee} + u_{ie} + u_{\rm vdW}.
\end{equation}
The interaction contribution to the internal energy consists of four terms:
\begin{itemize}
\item Ion-ion interaction contribution:
this term which, in general, yields the largest contribution is generated by
the OCP subsystem of the protons.
For the OCP energy of protons many expressions are available, e.g. \cite{Slattery80}.
We have used here the most precise formula due to Kahlbaum \cite{Kahlbaum96}
which interpolates between the Debye region,
$u_{ii} \sim \Gamma^{3/2}$, and the high density fluid, $u_{ii} \sim \Gamma$.

\item Electron-electron interaction: This term corresponds to the OCP energy of the electron subsystem.
We used the rather simple expressions used in earlier work \cite{Pade85,Pade90}.

\item Electron-proton interaction: This term corresponds to the interaction
between the two OCP subsystems which is mostly due to polarization effects.
Again, we used the rather simple expressions
proposed in earlier work \cite{Pade85,Pade90}.

\item Van der Waals contribution: In the region of densities and temperatures
defined above this contribution gives only a small correction. Therefore, this
term was approximated here in the simplest way
by a second virial contribution. The
neutral particles
were treated as hard spheres.
\end{itemize}

In the region of densities which are studied here, molecules do not play
a role, therefore, the formation of molecules was taken into account
only in a very rough approximation according to Ref.~\cite{Pade85}.
The number density of the neutrals was calculated on the basis
of a nonideal Saha equation. We
restricted this comparison to a region where the number density of
neutrals is
relatively small, the degree of ionization being
larger than $75 \%$.

The contributions to the pressure were calculated, in part,
from scaling relations e.g. we used
$p_{ii} = u_{ii} / 3$, and, for the other (smaller) contributions, by numerical
differentiation of the free energy given earlier \cite{Pade85,Pade90}.
In a similar way, the chemical potential which appears in the nonideal Saha
equation was obtained.
For the partition function in the Saha equation we used the Brillouin-Planck-Larkin
expression \cite{green-book,EbFo91}. The solution of the nonideal Saha
equation which determines the degree of ionization (the density of the atoms)
was solved by up to 100 iterations starting
from the ideal Saha equation.

\vspace{-0.5cm}
\begin{figure}[p] 
\centerline{ 
\psfig{file=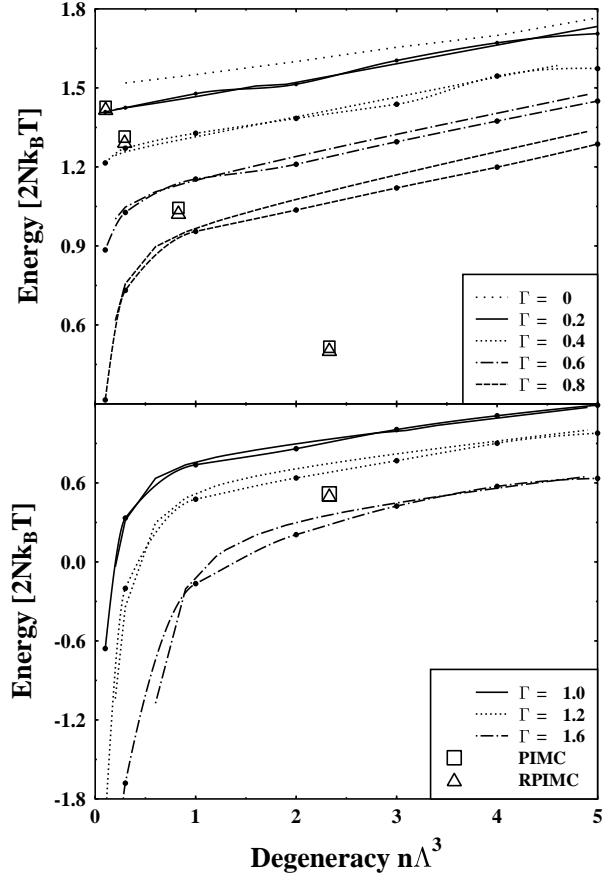,height=14cm}} 
\vspace{-1cm} 
\caption[]{\label{figeb1} Comparison of Pad\'e calculations (lines without 
symbols) for the internal 
energy with the direct PIMC results (lines with full circles). 
} 
\end{figure} 

Since all the expression described so far are given in analytic form, the
calculation of
about 1000 data points for energy and pressure takes less than a minute on a PC.
The result of our calculations for density-temperature points in the ``rhomb of moderate
nonideality and moderate degeneracy'' are given in Figs.~2,3.
Further, we give in TAB. I several data points obtained
from the Pad\'e formulas.
Since the Pad\'e formulas used here do not apply to low temperatures, we included
in TAB. I only Pad\'e data for $T > 10^5 K$.
\vspace{.0cm} 
\begin{figure}[p] 
\centerline{ 
\psfig{file=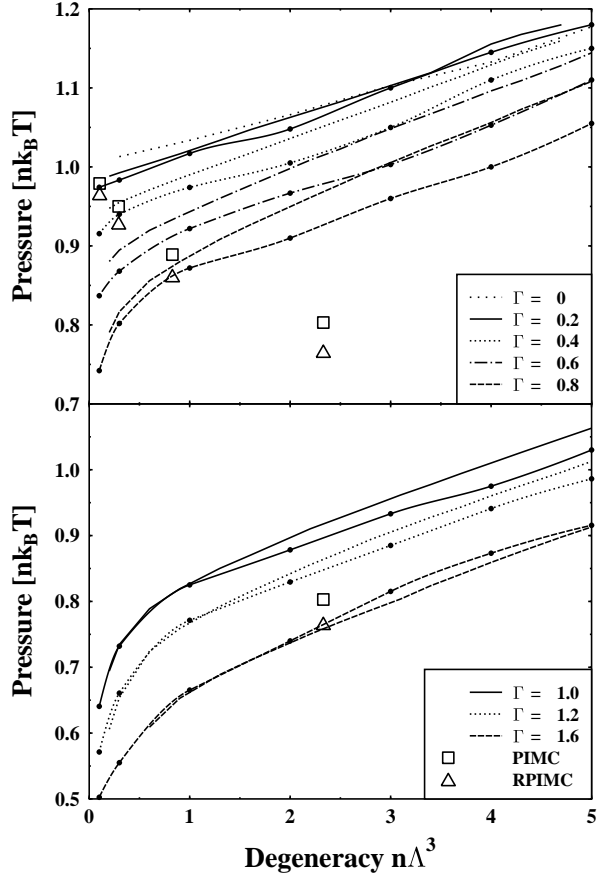,height=14cm}} 
\vspace{-0.5cm} 
\caption[]{\label{figeb2}Comparison of Pad\'e calculations (lines without 
symbols) of the pressure (in units of the Boltzmann pressure) with  direct 
PIMC simulation results (lines with full circles). 
} 
\end{figure} 

Summarizing the results for the internal energy and for the pressure we find
that the Pad\'e results,
with a few exceptions, agree well with the PIMC data in the
region of the density temperature plane, where $\Gamma \le 1.6$ and
$n\Lambda^3 \le 5$. The agreement is particularly good for the energies. [The
larger deviations for the pressure may be due to the numerical differentiation.]
In fact, the Pad\'e formulas used here in combination with the chemical picture
works only in the case that the plasma is strongly ionized, i.e. the degree of
ionization is larger than $75\%$.
The description of the region where a higher percentage of atoms and, due to this,
also molecules is present needs a more refined chemical picture
\cite{sbt95,BeEb99,MiBe00}.

\vspace{-2.5cm}
\begin{figure}[p] 
\centerline{ 
\psfig{file=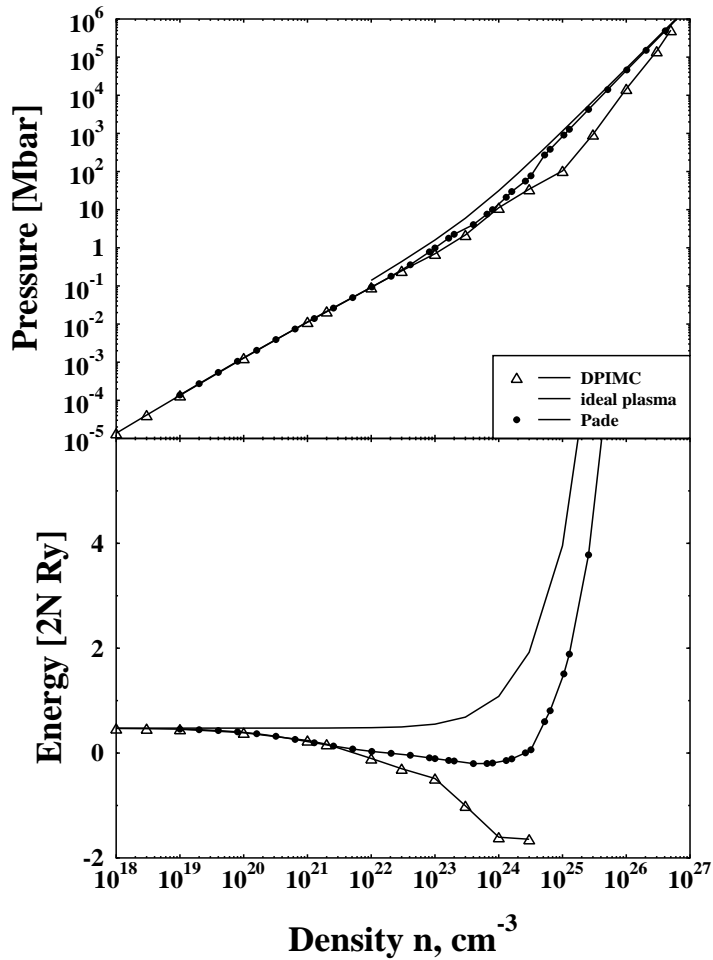,height=18cm,width=13cm}} 
\vspace{-1.5cm} 
\caption[]{\label{figeb3}Comparison of Pad\'e calculations (lines without 
symbols) of the pressure (in units of the Boltzmann pressure) with  direct 
PIMC simulation results (lines with full circles) for an isotherm $T=50,000K$. 
} 
\end{figure} 

Finally, we compare the Pad\'e and PIMC data along the isotherm $T = 50,000 K$ which 
is given in Fig.~5. This figure shows the transition from a classical ideal gas 
(low density) to a nearly ideal quantum gas (limit of high density). In the 
central part, $n\lesssim 10^{19} {\rm cm}^{-3} \lesssim 10^{25} {\rm cm}^{-3}$, 
Coulomb interaction 
leads to strong deviations from the behavior of an ideal plasma. The strong increase 
of the energy at high density is due to the Mott effect and to the increase of 
the ideal quantum contribution to the electron energy.
Comparing the Pad\'e and PIMC results, we find good agreement up to electron densities 
$n=10^{22}{\rm cm}^3$. For higher densities, the deviations are growing. For 
intermediate densities, $n\lesssim 10^{22} {\rm cm}^{-3} \lesssim 10^{24} {\rm cm}^{-3}$ 
the PIMC data are more reliable. On the other hand, in the limit of very high density, 
$r_s \ll 1$, the Pad\'e results are known to correctly approach the ideal quantum plasma 
limit whereas the PIMC data should be regarded as preliminary due to the extremely 
high electron degeneracy. Interestingly, we find that at high density the Pad\'e data 
approaches the ideal curves earlier than the PIMC data which is important for 
further improvement of the presented Monte Carlo approach.

\section{Discussion}

This work is devoted to the investigation of the thermodynamic properties of
hot dense partially ionized plasmas in the pressure range between 0.1 and 100 Mbar.
Most of the new results are based on a Quantum Monte Carlo study
of a correlated proton-electron system with degenerate electrons
and classical 
protons. In this paper, we gave a detailed derivation of improved estimators 
for the internal energy and the equation of state for use in direct fermionic 
path integral simulations. Also, we 
gave a rigorous justification for the use of an effective quantum pair potential 
(Kelbg potential) in PIMC simulations.

Further, we compared our direct PIMC results with independent
restricted PIMC data of Militzer and Ceperley for one isochor corresponding to
$r_s=1.86$, Fig. 2. We found very good quantitative agreement between the two PIMC 
methods for temperatures in the range of $50,000K\le T \le 10^6K$,
where $\Gamma \lesssim 3$ and $n_e \Lambda_e^3 \lesssim 10$. This region is
particularly complicated as here pressure and temperature ionization occur and,
therefore, an accurate and consistent treatment of scattering and bound states
is crucial. 
This agreement is remarkable because the two
simulation methods are completely
 independent 
and use essentially different approximations.
We, therefore, expect that the results for the thermodynamic
properties of high pressure hydrogen plasmas in this temperature-density range 
are reliable within the limits of
 the simulation accuracy. This is the main result
of the present paper.

In future work, it will be important to extend the range of agreement. To analyze 
the deviations between the simulation methods, we also included some data for $r_s=1.86$ 
and lower temperatures, $10,000K\le T \le 50,000K$, Fig.~2. At this point, no 
conclusive answer about the reasons of the deviations can be given. 
For these parameters, the electron degeneracy is growing rapidly and, therefore, each 
of the simulation methods is becoming less reliable. So these data should be regarded 
as preliminary results which will be useful for future improvements of the simulations.

Furthermore, the Monte Carlo results allowed us to develop and test analytical
approximations of Pad\'e-type which are improvements of earlier approximations 
\cite{green-book,Pade85,Pade90,EbFo91}
 in a region in the density-temperature plane
 bounded by
 $\Gamma \le 1.6$ and $n_e \Lambda_e^3 \le 5$. 
This is a region
of {\em moderate nonideality and degeneracy and high degree
of ionization}.
We have shown that for these parameters, the Pad\'e approximations which interpolate 
between the limits where
 theoretical results are available agree  well with the
Monte Carlo data, cf. Figs.~2-4 and Table I. Thus, these approximations provide a useful tool 
for the description of
 these plasmas which include 
hydrogen at a pressur between 0.1 and 100 Mbar. 
At lower temperature, deviations from the Monte Carlo data are growing, cf. Fig.~2. 
This is mostly due to the growing role of bound states.
Whether the Pad\'e approximations,
in combination with an improved chemical picture (mass action law), continue to work 
at lower
 temperatures, has still to be explored, first steps are under way \cite{MiBe00}.

Also, we showed some data for $T=50,000K$ and higher pressure, up to $p\sim 10^6$ Mbar, 
Fig.~5. Here the Monte Carlo simulations are particularly difficult due to the high
electron degeneracy, and they can benefit from the Pade simulations, as the latter 
correctly reproduce the high-density limit, $r_s \ll 1$. 

\section{Acknowledgements}
We acknowledge stimulating discussions with
W.D.~Kraeft, D.~Kremp and M.~Schlanges. We thank D.M.~Ceperley and
B.~Militzer for discussions on PIMC concepts and for
providing us with the data of Ref. \cite{militzer-etal.00} prior to
publication. Further we thank J.~Ortner for informing us about an alternative
derivation for the off-diagonal elements of the interaction potential.
This work was made possible by generous support from the Deutsche
Forschungsgemeinschaft (Mercator-Programm) for VSF and by a grant for
CPU time at the NIC J\"ulich.

\begin{table}
\caption{
Direct versus restricted PIMC [23] simulation results (upper and middle lines,
respectively) and results of Pad\'e calculations (numbers in the lowest lines)
for the pressure $p({\rm Mbar})$ and energy $E(2{\rm N Ry})$ for dense
hydrogen (deuterium [23]) for $r_s=1.86$}
\begin{tabular}{|r|c|c|c|c|}
$T, 1000K$ & $n\Lambda^3$ & $\Gamma$ & $p, {\rm Mbar}$ & $E, 2{\rm N Ry}$ \\
\hline\hline
1000 & 0.10  & 0.169  & 67.74 $\pm$ 0.02  & 9.050 $\pm$ 0.005    \\
&&&66.86 $\pm$ 0.08 & 9.018 $\pm$ 0.015\\
&&& 67.38     & 9.063 \\
\hline
500  & 0.29  & 0.339  & 32.85 $\pm$ 0.03   & 4.169 $\pm$ 0.003  \\
&&&32.13 $\pm$ 0.05 & 4.114 $\pm$ 0.007 \\
&&& 31.91     & 4.162 \\
\hline
250 & 0.83  & 0.679  & 15.37 $\pm$ 0.01  &  1.654 $\pm$ 0.005 \\
&&&14.91 $\pm$ 0.03&  1.629 $\pm$ 0.007\\
&&& 14.40     & 1.679 \\
\hline
125  & 2.33  & 1.350  & 6.98 $\pm$ 0.01  & 0.412 $\pm$ 0.005   \\
&&&6.66 $\pm$ 0.02& 0.404 $\pm$ 0.004 \\
&&& 6.47   & 0.471 \\
\hline
62.5 & 6.58  & 2.701  & 3.07 $\pm$ 0.02   & -0.248 $\pm$ 0.005  \\
&&&2.99 $\pm$ 0.04 & -0.140 $\pm$ 0.007\\
\hline
31.25  & 18.48  & 5.376  & 2.20 $\pm$ 0.01    & -2.377 $\pm$ 0.005   \\
&&&1.58 $\pm$ 0.07 & -0.360 $\pm$ 0.010\\
\end{tabular}
\end{table}


\begin{references}
\bibitem{imada} The method of Imada \cite{imada84}
ignores the sign of the exchange determinant, cf. Eq. (\ref{psi}).
We have found that Imada's approximation
leads to essential errors for strongly degenerate systems.
We underline that, in contrast, our method rigorously takes into account
the sign of the determinant of each Monte Carlo configuration in
the calculation of thermodynamic averages, such as pressure and
energy, Eqs. (\ref{eos}, \ref{energy}).
\end{references}
\end{document}